\documentclass[12pt]{article}
\usepackage{graphicx}
\begin{document}

\begin{center}

{\Large \bf
Einstein's $E = mc^2$ derivable  \\[1.0ex]
from Heisenberg's Uncertainty Relations }

\vspace{16mm}

Sibel Ba{\c s}kal  \\
Department of Physics, Middle East Technical University, 06800 Ankara, Turkey \\[2ex]

 Young S. Kim \\
Center for Fundamental Physics, University of Maryland College Park,\\
Maryland,  20742, USA \\[2ex]

 Marilyn E. Noz \\
Department of Radiology, New York University, New York, NY 10016, USA

\end{center}

\vspace{2mm}

\abstract{Heisenberg's uncertainty relation can be written in terms of the
 step-up and step-down operators in the harmonic oscillator representation.
 It is noted that the single-variable Heisenberg commutation relation
 contains the symmetry of the $Sp(2)$ group which is isomorphic to the
 Lorentz group applicable to one time-like dimension and two space-like
 dimensions, known as the $O(2,1)$ group.  This group has three independent
 generators.  The one-dimensional step-up and step-down operators can be
 combined into one two-by-two Hermitian matrix which contains three
 independent operators.  If we use a two-variable Heisenberg commutation
 relation, the two pairs of independent step-up, step-down operators can
 be combined into a four-by-four block-diagonal Hermitian matrix with six
 independent parameters.  It is then possible to add one off-diagonal
 two-by-two matrix and its Hermitian conjugate to complete the four-by-four
  Hermitian matrix.  This off-diagonal matrix has four independent
 generators.  There are thus ten independent generators.
 It is then shown that these ten generators can be linearly combined to
 the ten generators for the Dirac's two oscillator system leading
 to the group isomorphic to the de Sitter group $O(3,2)$, which can then
 be contracted to the inhomogeneous Lorentz group with four translation
 generators corresponding to the four-momentum in the Lorentz-covariant world.
 This Lorentz-covariant four-momentum is known as Einstein's $E = mc^{2}.$}

\vspace{3cm}

\noindent Based on an invited paper presented by Young S. Kim
at the 16th International Conference on Squeezed States and Uncertainty Relations
(Madrid, Spain, 2019); published in the Quantum Reports, Vol. 1 (2),
236 - 251 (2019), https://www.mdpi.com/2624-960X/1/2/21.

\newpage

\section{Introduction}
Let us start with Heisenberg's commutation relations
\begin{equation}\label{com01}
\left[x_{i},P_{j}\right] = i~\delta_{ij} ,
\end{equation}
with
\begin{equation}\label{p01}
P_{i} = -i\frac{\partial}{\partial x_{i}},
\end{equation}
where $i = 1, 2, 3,$ correspond to the $x, y, z$ coordinates respectively.

With these $x_{i}$ and $P_{i}$, we can construct the following three  operators,
\begin{equation}\label{j01}
J_{i} = \epsilon_{ijk}x_{j}P_{k} .
\end{equation}
These three operators satisfy the closed set of commutation relations:
\begin{equation}\label{com03}
   \left[J_{i}, J_{j} \right] = i\epsilon_{ijk}J_{k} .
\end{equation}
These $J_{i}$ operators generate rotations in the three-dimensional
space.  In mathematics, this set is called the Lie algebra of the
rotation group.  This is a direct consequence of Heisenberg's commutation
relations.

In quantum mechanics, each $J_{i}$ corresponds to the angular momentum
along the $i$ direction.  The remarkable fact is that it is also possible
to construct the same Lie algebra with two-by-two matrices.  These
matrices are of course the Pauli spin matrices, leading to the observable
angular momentum not seen in classical mechanics.

As the expression shows in Eq.(\ref{p01}), each $P_{i}$ generates a
translation along the $i^{th}$ direction.  Thus, the three translation
generators, together with the three rotation generators constitute
the Lie algebra of the Galilei group, with the additional commutation
relations:
\begin{equation}\label{com05}
\left[J_{i}, P_{j}\right] = i\epsilon_{ijk}P_{k} .
\end{equation}
This set of commutation relations together with those of Eq.(\ref{com03})
constitute a closed set for both $P_{i}$ and $J_{i}$.
This set is called the Lie algebra of the Galilei group.  This group is
the basic symmetry group for the Schr\"odinger or non-relativistic
quantum mechanics.

In the Schr\"odinger picture, the generator $P_{i} $ corresponds to
the particle momentum along the $i$ direction.  In addition, the time
translation operator is
\begin{equation}
P_{0} = i\frac{\partial}{\partial t} .
\end{equation}
This corresponds to the energy variable.

Let us go to the Lorentzian world. Here we have to take into account
the generators of the boosts. The generators thus include the time
variable, and the generator of boosts along the $i$ direction is
\begin{equation}\label{b03}
K_{i} = i\left(x_{i}\frac{\partial}{\partial t}
      + t\frac{\partial}{\partial x_{i}} \right) .
\end{equation}
These generators satisfy the commutation relations
\begin{equation}\label{com07}
   \left[K_{i}, K_{j}\right] = -i\epsilon_{ijk} J_{k} .
\end{equation}
Thus, these three boost generators alone cannot constitute a
closed set of commutation relations (Lie algebra).

With $J_{i}$, these boost generators satisfy
\begin{equation} \label{com09}
 \left[J_{i}, K_{j}\right] = i~\epsilon_{ijk} K_{k} .
\end{equation}
With $P_{i}$, they satisfy the relation
\begin{equation}\label{com11}
\left[P_{i}, K_{i}\right] = i\delta_{0i} P_{0} .
\end{equation}
Thus, the commutation relations of
Eqs.(\ref{com03},\ref{com05}, \ref{com07},\ref{com09},\ref{com11})
constitute a closed set of the ten generators.  This closed set is
commonly called the Lie algebra of the Poincar\'e symmetry.

The three rotation and three translation generators are contained in
or derivable from Heisenberg's commutation relations, and the time
translation operator is seen in the Schr\"odinger equation.  They are
all Hermitian operators corresponding to dynamical variables.
On the other hand, the three boost generators of Eq.(\ref{b03})
are not derivable from the Heisenberg relations.  Furthermore,
they do not appear to correspond to observable quantities~\cite{dir49}.

The purpose of this paper is to show that the Lie algebra of the Poincar\'e
symmetry is derivable from the Heisenberg commutation relations.  For this
purpose, we first examine the symmetry of the Heisenberg commutation
relation using the Wigner function in the phase space.  It is noted
that the single-variable relation contains the symmetry of the
Lorentz group applicable to two space-like and one time-like dimensions.

\begin{figure}%[thb]
\centerline{\includegraphics[width=16cm]{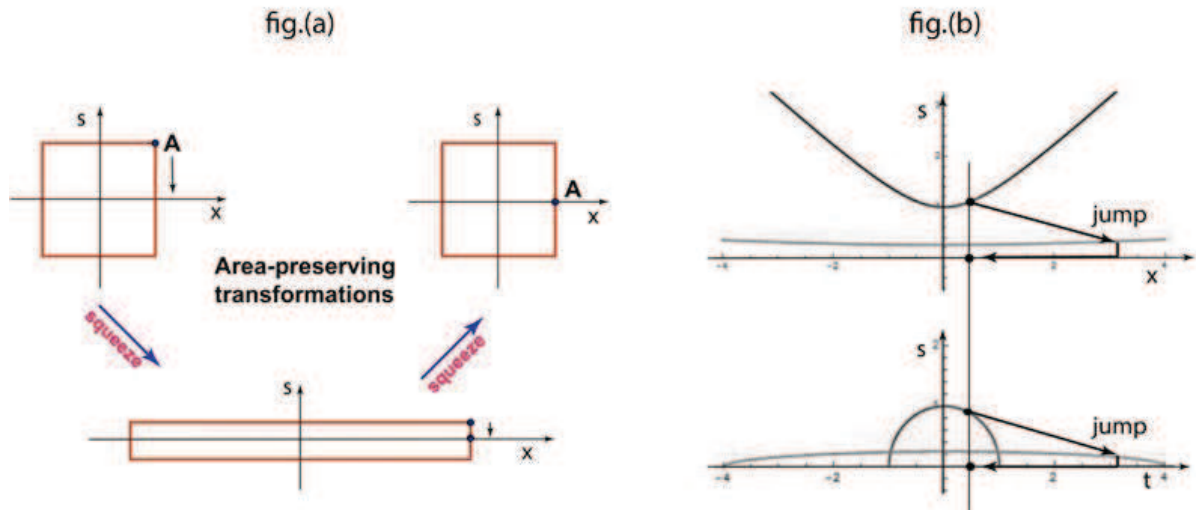}}
\caption{The In{\"o}n{\"u}-Wigner contraction procedure interpreted as
squeeze transformations.  In fig.(a), the square becomes a narrow
rectangle during the squeeze process.  When the rectangle becomes narrow
enough, the point A can be moved to the horizontal axis.  Then, the inverse
squeeze brings back the rectangle to the original shape.  The point A remains
on the horizontal axis.
In fig.(b), both the hyperbola and the circle become flattened  to the
horizontal axis, during the initial squeeze. The point on the curve moves
to the horizontal axis.  This point moves back to its finite position
during the inverse squeeze.}\label{contrac77}
\end{figure}

As Dirac noted in 1963~\cite{dir63}, two coupled oscillators
lead to the symmetry of the $O(3,2)$ or the Lorentz group applicable
to the three space-like directions and two time-like directions.
As is illustrated in Fig.~\ref{contrac77}, it is possible to contract
one of those two time variables of this $O(3,2)$ group into
the inhomogeneous Lorentz group consisting of the Lorentz group
applicable to the three space-like dimensions and one time-like
direction, plus four translation generators corresponding to
the energy-momentum four-vector.  This of course leads to
Einstein's energy-momentum relation of $E = mc^{2}$.

%----------------------------------------------------------------------
\begin{figure}%[thb]
\centerline{\includegraphics[width=11cm]{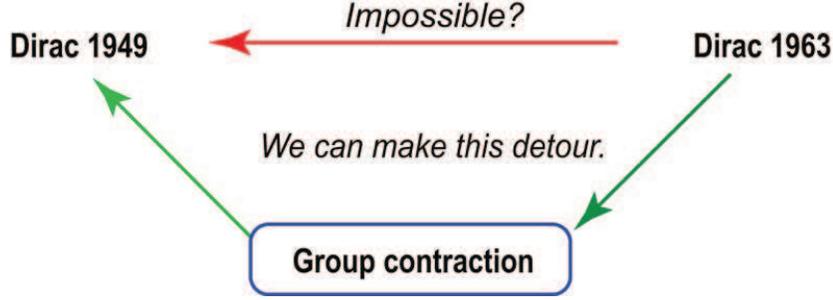}}
\caption{According to Dirac's 1949 paper, the task of constructing
quantum mechanics is essentially constructing a representation of the
inhomogeneous Lorentz group.  In his 1963 paper, Dirac constructed
the Lie algebra of the $O(3,2)$ de Sitter group from the algebra of
two harmonic oscillators, which is a direct consequence of Heisenberg's
uncertainty commutation relations. It is possible to derive the Lie
algebra of the inhomogeneous Lorentz group from that of $O(3,2)$
using the group-contraction procedure of In{\"o}n{\"u} and
Wigner~\cite{inonu53}.}\label{dirwig}
\end{figure}
%----------------------------------------------------------------------

In Sec.~\ref{single}, it is noted that the best way to study the symmetry
of the Heisenberg commutation relation is to use the Wigner function
for the Gaussian function for the oscillator state.  In the Wigner
phase space, this function contains the symmetry for the Lorentz group
applicable to two space-like dimensions and one time-like dimension.
This group has three generators.  This operation is equivalent
to constructing a two-by-two block-diagonal Hermitian matrix with
quadratic forms of the step-up and step-down operators.

In Sec.~\ref{2osc}, we consider two oscillators.
If these oscillators are independent, it is possible to construct
a four-by-four block diagonal matrix, where each block consists of
the two-by-two matrix for each operator defined in Sec.~\ref{single}.
Since the oscillators are uncoupled, this four-by-four
block-diagonal Hermitian matrix contains six independent generators.

If the oscillators are coupled, then to keep the overall four-by-four
block-diagonal matrix Hermitian, we need one
off-diagonal block matrix, with four independent quadratic forms. Thus,
the overall four-by-four matrix contains ten independent quadratic forms
of the creation and annihilation operators.

It is shown that these ten independent generators can be linearly
combined into the ten generators constructed by Dirac for the
the Lorentz-group applicable to three space-like dimensions and two
time-like dimensions, commonly called $O(3,2)$ group.

In Sec.~\ref{contrac}, using the boosts belonging to one of its
time-like dimensions, we contract
$O(3,2)$ to produce the Lorentz group applicable to one time dimension
and four translations leading to the four-momentum.  This
Lorentz-covariant four-momentum is commonly known as Einstein's
$E = mc^{2}$.

This paper is basically based on Dirac's paper  published in 1949 and
1963~\cite{dir49,dir63}.  As is illustrated in Fig.\ref{dirwig},
we show here that the space-time symmetry of quantum mechanics mentioned
in his 1949 paper is derivable from his two-oscillator system discussed
in 1963.   The route is the group contraction procedure of
In{\"o}n{\"u} and Wigner~\cite{inonu53}.

Indeed, Dirac made his lifelong efforts to synthesize quantum
mechanics and special relativity from 1927~\cite{dir27}.  In and before
1949, he treated quantum mechanics and special relativity as two
separate scientific disciplines, and he then attempted to synthesized
them. Thus, it is of interest to see how Dirac's idea evolved during
the period 1929-49. We shall give a brief review of Dirac's
efforts during the period in the Appendix.

%%%%%%%%%%%%%%%%%%%%%%%%%%%%%%%%%%%%%%%%%%%%%%%%%%%%%%%%%%%%%%%%%%%%%%%%%%%%%%%%%%%%%%%%

\section{Symmetries of the Single-mode States}\label{single}
Heisenberg's uncertainty relation for a single Cartesian variable takes the form
\begin{equation}\label{102}
   [x, p] = i .
\end{equation}
with
$$
p = -i\frac{\partial}{\partial x} .
$$
Very often, it is more convenient to use the operators
\begin{equation}
 a = \frac{1}{\sqrt{2}}(x + ip), \qquad  a^{\dag} = \frac{1}{\sqrt{2}}(x - ip)
\end {equation}
with
\begin{equation}
\left[a, a^{\dag}\right] = 1 .
\end{equation}
This aspect is well known.

The representation based on  $a$ and $a^{\dag}$ is known as the harmonic oscillator
representation of the uncertainty relation and is the basic language for the Fock
space for particle numbers.  This representation is therefore the basic language
for quantum optics.

Let us next consider the quadratic forms: $aa, a^{\dag}a^{\dag}, aa^{\dag}$,
and $ a^{\dag}a$.  Then the linear combination
\begin{equation}
aa^{\dag} - a^{\dag} a = 1,
\end{equation}
according to the uncertainty relation.  Thus, there are three independent
quadratic forms, and we are led to the following two-by-two  matrix:
\begin{equation}\label{m01}
\pmatrix{ \left(aa^{\dag} + a^{\dag}a\right)/2   & aa \cr
   a^{\dag}a^{\dag}           &   \left(aa^{\dag} +
   a^{\dag}a\right)/2 } .
 \end{equation}
This matrix leads to the following three independent operators:
\begin{equation}
J_{2} = \frac{1}{2} \left(aa^{\dag} + a^{\dag}a\right),  \quad
K_{1} = \frac{1}{2} \left(a^{\dag}a^{\dag} + a a\right), \quad
K_{3} = \frac{i}{2}  \left(a^{\dag}a^{\dag} - a a\right).
\end{equation}
They produce the following set of closed commutation relations:
\begin{equation}\label{115}
 \left[J_{2}, K_{1}\right] = - iK_{3},  \qquad
 \left[J_{2}, K_{3}\right] = iK_{1},   \qquad
 \left[K_{1}, K_{3}\right] = iJ_{2}.
\end{equation}
This set is commonly called the Lie algebra of the $Sp(2)$ group,
locally isomorphic to the Lorentz group applicable to one time
and two space coordinates.

\begin{figure}[thb]
\centerline{\includegraphics[width=8cm]{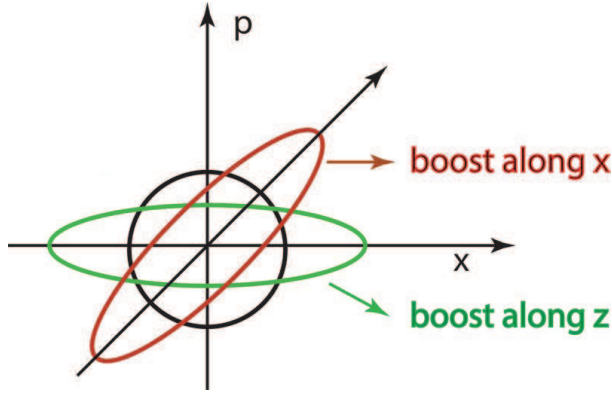}}
\caption{Rotations and squeezes in the phase space produced by the
$Sp(2)$ transformations.  The squeeze along the $x$ direction
corresponds to the Lorentz boost along the $z$ direction, while
the squeeze along the $45^o$ degree corresponds to the boost along
the $x$ direction.  The rotation by $45^o$ corresponds to the
rotation by $90^o$ around the $y$ axis.}\label{sp2}
\end{figure}

The best way to study the symmetry property of these operators is to use
the Wigner function for the ground-state oscillator which
takes the form~\cite{hkn88,kiwi90ajp,knp91,dodo03}
\begin{equation}\label{103}
W(x,p) = \frac{1}{\pi} \exp\left[-\left(x^2 + p^2\right)\right].
\end{equation}
This distribution is concentrated in the circular region around the origin.
Let us define the circle as
\begin{equation}\label{105}
x^{2} + p^{2} = 1.
\end {equation}
We can use the area of this circle in the phase space of $x$ and $p$ as the
minimum uncertainty.  This uncertainty is preserved under rotations in the
phase space and also under squeezing. These transformations can be written as
\begin{equation}\label{107}
\pmatrix{\cos\theta & -\sin\theta \cr \sin\theta & \cos\theta }
\pmatrix{x \cr p }, \qquad
\pmatrix{e^{\eta} &  0 \cr 0 &  e^{-\eta}} \pmatrix{x \cr p },
\end{equation}
respectively.
The rotation and the squeeze are generated by
\begin{equation}\label{109}
J_{2} = - i\left(x \frac{\partial}{\partial p} - p\frac{\partial}{\partial x } \right),
\qquad
K_{1} = -i \left(x\frac{\partial}{\partial x} - p \frac{\partial}{\partial p} \right).
\end{equation}
If we take the commutation relation with these two operators, the result is
\begin{equation}\label{111}
\left[J_{2}, K_{1}\right] = -i K_{3},
\end{equation}
with
\begin{equation}\label{113}
K_{3} = -i \left(x\frac{\partial}{\partial p} + p \frac{\partial}{\partial x} \right).
\end{equation}
Indeed, as before, these three generators form the closed set of commutation
which form the Lie algebra of the $Sp(2)$ group, isomorphic
to the Lorentz group applicable to two space and one time dimensions.
This isomorphic correspondence is illustrated in Fig.~\ref{sp2}.

%----------------------------------------------------------------------
%----------------------------------------------------------------------

Let us consider the Minkowski space of $(x, y, z, t)$.  It is possible to
write three four-by-four matrices satisfying the Lie algebra of Eq.(\ref{115}).
The three four-by-four matrices satisfying this set
of commutation relations are:
\begin{equation}\label{119}
J_{2} = \pmatrix{0 & 0 & i & 0 \cr 0 & 0 & 0 & 0 \cr i & 0 & 0 & 0 \cr
    0 & 0 & 0 & 0 },\quad
K_{1} = \pmatrix{0 & 0 & 0 & i \cr 0 & 0 & 0 & 0 \cr 0 & 0 & 0 & 0 \cr
    i & 0 & 0 & 0 },\quad
K_{3} = \pmatrix{0 & 0 & 0 & 0 \cr 0 & 0 & 0 & 0 \cr 0 & 0 & 0 & i \cr
    0 & 0 & i & 0 }.
\end{equation}
However, these matrices have null second rows and null second columns.
Thus, they can generate Lorentz transformations applicable only to the
three-dimensional space of $(x,z,t)$, while the $y$ variable remains invariant.
Thus, this single-oscillator system cannot describe what happens in the full
four-dimensional Minkowski space.

Yet, it is interesting, the oscillator system can produce three different
representations sharing the same Lie algebra with the (2 + 1)-dimensional
Lorentz group, as shown in Table~\ref{tab11}.

\begin{table}%[thb]
\caption{Transformation for the Gaussian function, in terms of harmonic oscillators,
 two-dimensional phase space, and the four-dimensional Minkowski space.}\label{tab11}
\vspace{0.5mm}
\begin{center}
\begin{tabular}{ccccccccc}
\hline
\hline\\[-0.4ex]
\hspace{1mm}& Generators &\hspace{1mm} & Oscillator &\hspace{1mm}& Phase space   &\hspace{1mm}&  Lorentz
\\[0.8ex]
\hline\\ [-1.0ex]
\hspace{1mm}&
$J_{2}$
&\hspace{1mm} &  $ {1\over 2}\left(a a^{\dag} + a^{\dag}a\right) $
                  & \hspace{1mm}&
$\frac{1}{2}\sigma_{2}$
 & \hspace{1mm}&
$\pmatrix{0 & 0 & i & 0 \cr 0 & 0 & 0 & 0 \cr i & 0 & 0 & 0 \cr
    0 & 0 & 0 & 0 }$
\\[6ex]
\hline\\ [-0.7ex]
\hspace{1mm}&
$K_{1} $
&\hspace{1mm} &
$  {1\over 2i}\left(a^{\dag}a^{\dag} +  aa\right)$
                & \hspace{1mm} &
$ \frac{i}{2}\sigma_{1} $  & \hspace{1mm} &
$ \pmatrix{0 & 0 & 0 & i \cr 0 & 0 & 0 & 0 \cr 0 & 0 & 0 & 0 \cr
    i & 0 & 0 & 0 }$
\\[6.0ex]
\hline\\ [-0.7ex]
\hspace{1mm}& $K_{3} $ &\hspace{1mm} &  $
{1\over 2} \left(a^{\dag}a^{\dag} - aa \right),
                  $ & \hspace{1mm}&
$\frac{i}{2}\sigma_{3}$
  & \hspace{1mm}&
$ \pmatrix{0 & 0 & 0 & 0 \cr 0 & 0 & 0 & 0 \cr 0 & 0 & 0 & i \cr
    0 & 0 & i & 0 } $
\\[6.0ex]
%----------------------------------------------
\hline
\hline\\[-0.4ex]
\end{tabular}
\end{center}
\end{table}
%----------------------------------------------------------------------------------

%%%%%%%%%%%%%%%%%%%%%%%%%%%%%%%%%%%%%%%%%%%%%%%%%%%%%%%%%%%%%%%%%%%%%%%%%%%%%%%%%%%%%%%%

\section{Symmetries from Two Oscillators}\label{2osc}

In order to generate Lorentz transformations applicable to the full
Minkowskian space, we may need two Heisenberg commutation relations.
Indeed, Paul A. M. Dirac started this program in 1963~\cite{dir63}.
It is possible to write the two uncertainty relations using two harmonic
oscillators as
\begin{equation}\label{121}
\left[a_{i}, a^{\dag}_{j}\right] = \delta_{ij} ,
\end{equation}
with
\begin{equation}\label{123}
  a_{i} = \frac{1}{\sqrt{2}}\left(x_{i} + ip_{i} \right), \qquad
  a^{\dag}_{i} = \frac{1}{\sqrt{2}}\left(x_{i} - ip_{i} \right),
\end{equation}
and
\begin{equation}\label{125}
  x_{i} = \frac{1}{\sqrt{2}}\left(a_{i} + a^{\dag}_{i} \right), \qquad
  p_{i} = \frac{i}{\sqrt{2}}\left(a^{\dag}_{i} - a_{i} \right),
\end{equation}
 where $i$ and $j$ could be 1 or 2.

As in the case of the two-by-two matrix given in Eq.~\ref{m01}, we can consider
the following four-by-four block-diagonal matrix if the oscillators are not
coupled:
\begin{equation}\label{m03}
\pmatrix{\left(a_{1}a_{1}^{\dag} + a_{1}^{\dag}a_{1}\right)/2   & a_{1}a_{1} & 0 & 0  \cr
a_{1}^{\dag}a_{1}^{\dag} & \left(a_{1}a_{1}^{\dag} + a_{1}^{\dag}a_{1}\right)/2 & 0 & 0 \cr
0 & 0 & \left(a_{2}a_{2}^{\dag} + a_{2}^{\dag}a_{2}\right)/2 & a_{2}a_{2}   \cr
0 & 0 & a^{\dag}_{2}a_{2}^{\dag} & \left(a_{2}a^{\dag}_{2}+ a^{\dag}_{2} a_{2}\right)/2 }.
\end{equation}
There are six generators in this matrix.

We are now interested in coupling them by filling in the off-diagonal blocks.  The
most general forms for this block are the following two-by-two matrix and its Hermitian
conjugate:
 \begin{equation}\label{m05}
\pmatrix{a^{\dag}_{1}a_{2} & a_{1}a_{2}  \cr  {} & {} \cr
a^{\dag}_{1}a^{\dag}_{2} & a_{1}a^{\dag}_{2}}
\end{equation}
with four independent generators.  This leads to the following four-by-four
matrix with ten (6~+~4) generators:
 \begin{equation} \label{m07}
\pmatrix{\left(a_{1}a_{1}^{\dag} + a^{\dag}_{1}a_{1}\right)/2   & a_{1}a_{1} &
a^{\dag}_{1}a_{2} & a_{1}a_{2}  \cr
a_{1}^{\dag}a_{1}^{\dag} & \left(a_{1}a_{1}^{\dag} + a_{1}^{\dag}a_{1}\right)/2 &
a^{\dag}_{1}a^{\dag}_{2} & a_{1}a^{\dag}_{2} \cr
a_{1}a^{\dag}_{2} & a_{1}a_{2} & \left(a_{2}a_{2}^{\dag} + a_{2}^{\dag}a_{2}\right)/2 &
a_{2}a_{2}   \cr
a^{\dag}_{1}a^{\dag}_{2} & a^{\dag}_{1}a_{2} & a^{\dag}_{2}a_{2}^{\dag} &
\left(a_{2}a^{\dag}_{2}+ a^{\dag}_{2} a_{2}\right)/2 }.
\end{equation}

With these ten elements, we can now construct the following four rotation-like
generators:

\begin{eqnarray}\label{303}
&{}& J_{1} = {1\over 2}\left(a^{\dag}_{1}a_{2} + a^{\dag}_{2}a_{1}\right) ,
\qquad
 J_{2} = {1\over 2i}\left(a^{\dag}_{1}a_{2} - a^{\dag}_{2}a_{1}\right), \nonumber \\[1ex]
&{}& J_{3} = {1\over 2} \left(a^{\dag}_{1}a_{1} - a^{\dag}_{2}a_{2} \right),
\qquad
S_{0} = {1\over 2}\left(a^{\dag}_{1}a_{1} + a_{2}a^{\dag}_{2}\right) ,
\end{eqnarray}

and six squeeze-like generators:
\begin{eqnarray}\label{304}
&{}&  K_{1} = -{1\over 4}\left(a^{\dag}_{1}a^{\dag}_{1} + a_{1}a_{1} -
                      a^{\dag}_{2}a^{\dag}_{2} - a_{2}a_{2}\right) ,   \nonumber \\[1ex]
&{}&  K_{2} = +{i\over 4}\left(a^{\dag}_{1}a^{\dag}_{1} - a_{1}a_{1} +
                       a^{\dag}_{2}a^{\dag}_{2} - a_{2}a_{2}\right) ,   \nonumber \\[1ex]
&{}&  K_{3} = +{1\over 2}\left(a^{\dag}_{1}a^{\dag}_{2} + a_{1}a_{2}\right) ,
\end{eqnarray}
and

\begin{eqnarray}\label{305}
&{}& Q_{1} = -{i\over 4}\left(a^{\dag}_{1}a^{\dag}_{1} - a_{1}a_{1} -
  a^{\dag}_{2}a^{\dag}_{2} + a_{2}a_{2} \right) ,   \nonumber \\[1ex]
&{}& Q_{2} = -{1\over 4}\left(a^{\dag}_{1}a^{\dag}_{1} + a_{1}a_{1} +
   a^{\dag}_{2}a^{\dag}_{2} + a_{2}a_{2} \right) ,\nonumber \\[1ex]
&{}& Q_{3} = +\frac{i}{2}\left(a_{1}^{\dag}a_{2}^{\dag} - a_{1}a_{2}\right).
\end{eqnarray}

There are now ten operators from Eqs.(\ref{303},\ref{304},\ref{305}), and
they satisfy the following Lie algebra as was noted by Dirac in 1963~\cite{dir63}:
\begin{eqnarray}\label{313}
&{}& [J_{i}, J_{j}] = i\epsilon _{ijk} J_{k} ,\quad
[J_{i}, K_{j}] = i\epsilon_{ijk} K_{k} , \nonumber\\[1ex]
&{}&[J_{i}, Q_{j}] = i\epsilon_{ijk} Q_{k} , \quad
[K_{i}, K_{j}] = [Q_{i}, Q_{j}] = -i\epsilon _{ijk} J_{k} ,  \nonumber\\[1ex]
&{}& [K_{i}, Q_{j}] = -i\delta_{ij} S_{0} , \quad [J_{i}, S_{0}] = 0 ,
\quad [K_{i}, S_{0}] =  -iQ_{i}, \quad [Q_{i}, S_{0}] = iK_{i} .
\end{eqnarray}
Dirac noted that this set is the same as the Lie
algebra for the $O(3,2)$ de Sitter group, with ten generators. This
is the Lorentz group applicable to the three-dimensional space with
two time variables.  This group plays a very important role in
space-time symmetries.

In the same paper, Dirac pointed out that this set of commutation
relations serves as the Lie algebra for the four-dimensional
symplectic group commonly called $Sp(4)$.
For a dynamical system consisting of two pairs of canonical variables
$x_{1}, p_{1}$ and $x_{2}, p_{2}$, we can use the four-dimensional
phase space with the coordinate variables defined as~\cite{hkn95jmp}
\begin{equation}\label{317}
\left(x_{1}, p_{1}, x_{2}, p_{2} \right).
\end{equation}
Then the four-by-four transformation matrix $M$ applicable to this
four-component vector is canonical if~\cite{abra78,goldstein80}
\begin{equation}\label{319}
M J \tilde{M} = J ,
\end{equation}
where $\tilde{M}$ is the transpose of the $M$ matrix,
with
\begin{equation}\label{321}
J = \pmatrix{0 & 1 & 0 & 0 \cr -1 & 0 & 0 & 0 \cr
             0 & 0 & 0 & 1 \cr 0 & 0 & -1 & 0 },
\end{equation}
which we can write in the block-diagonal form as
\begin{equation}
J = i\pmatrix{I & 0 \cr 0 & I}\sigma_{2},
\end{equation}
where $I$ is the unit two-by-two matrix.

According to this form of the $J$ matrix, the area of the phase space for
the $x_{1}$ and $p_{1}$ variables remains invariant, and the story is the
same for the phase space of $x_{2}$ and $p_{2}.$

%-------------------------------------------------------------------------------------
\begin{table}%[thb]
\caption{Transformation generators for the two-oscillator system.}\label{tab22}
\vspace{0.5mm}
\begin{center}
\begin{tabular}{cccccc}
\hline
\hline\\[-0.4ex]
\hspace{1mm}& Generators &\hspace{1mm} & Two Oscillators &\hspace{1mm}& Phase space \\[0.8ex]
\hline\\ [-1.0ex]
\hspace{1mm}&
$J_{1}$
&\hspace{1mm} &  $ {1\over 2}\left(a^{\dag}_{1}a_{2} + a^{\dag}_{2}a_{1}\right) $
                  & \hspace{1mm}&
$-\frac{1}{2}\pmatrix{0 & I\cr I & 0}\sigma_{2}$
\\[2.5ex]
\hline\\ [-0.7ex] \hspace{1mm}& $J_{2} $
&\hspace{1mm} &
$  {1\over 2i}\left(a^{\dag}_{1}a_{2} - a^{\dag}_{2}a_{1}\right)$
                & \hspace{1mm} &
$ \frac{i}{2} \pmatrix{0 & -I \cr I & 0} I $
\\[2.5ex]
\hline\\ [-0.7ex]
\hspace{1mm}&
$J_{3} $
&\hspace{1mm} &  $
{1\over 2} \left(a^{\dag}_{1}a_{1} - a^{\dag}_{2}a_{2} \right),
                  $ & \hspace{1mm}&
$\frac{1}{2}\pmatrix{-I & 0 \cr 0 & I}\sigma_{2}$
\\[2.5ex]
\hline\\ [-0.7ex]
\hspace{1mm}&
$S_{0}$
&\hspace{1mm} &  ${1\over 2}\left(a^{\dag}_{1}a_{1} + a_{2}a^{\dag}_{2}\right) ,
       $ & \hspace{1mm}&
$ \frac{1}{2}\pmatrix{I   & 0\cr 0 & I} \sigma_{2}$
\\[2.5ex]
%------------------------------------------------------------------------------
\hline\\ [-0.7ex]
\hspace{1mm}&
$K_{1}$
&\hspace{1mm} &  $
 -{1\over 4}\left(a^{\dag}_{1}a^{\dag}_{1} + a_{1}a_{1} -
                      a^{\dag}_{2}a^{\dag}_{2} - a_{2}a_{2}\right)$
                      & \hspace{1mm}&
$ \frac{i}{2}\pmatrix{I  & 0 \cr 0 & -I} \sigma_{1} $
\\[2.5ex]
%----------------------------------------------
\hline\\ [-0.7ex]
\hspace{1mm}&
$K_{2}$
&\hspace{1mm} &  $ +{i\over 4}\left(a^{\dag}_{1}a^{\dag}_{1}
      - a_{1}a_{1} + a^{\dag}_{2}a^{\dag}_{2} - a_{2}a_{2}\right)
       $ & \hspace{1mm}&
$ \frac{i}{2} \pmatrix{I & 0 \cr 0 & I} \sigma_{3}$
\\[2.5ex]
%----------------------------------------------
\hline\\ [-0.7ex]
\hspace{1mm}&
$K_{3}$
&\hspace{1mm} &
$ {1\over 2}\left(a^{\dag}_{1}a^{\dag}_{2} + a_{1}a_{2}\right)$
  & \hspace{1mm}&
$ -\frac{i}{2}\pmatrix{0 & I \cr I & 0}  \sigma_{1}$
\\[2.5ex]
%----------------------------------------------
\hline\\ [-0.7ex]
\hspace{1mm}&
$Q_{1}$
&\hspace{1mm} &  $
-{i\over 4}\left(a^{\dag}_{1}a^{\dag}_{1} - a_{1}a_{1} -
  a^{\dag}_{2}a^{\dag}_{2} + a_{2}a_{2} \right)
       $ & \hspace{1mm}&
$-\frac{i}{2}\pmatrix{I & 0 \cr 0 & -I }\sigma_{3}$
\\[2.5ex]
%----------------------------------------------
\hline\\ [-0.7ex]
\hspace{1mm}&
$Q_{2}$
&\hspace{1mm} &
 $ -{1\over 4}\left(a^{\dag}_{1}a^{\dag}_{1} + a_{1}a_{1} +
   a^{\dag}_{2}a^{\dag}_{2} + a_{2}a_{2} \right)
       $ & \hspace{1mm}&
$ \frac{i}{2}\pmatrix{I & 0 \cr 0 & I}\sigma_{1}$
\\[2.5ex]
%----------------------------------------------
\hline\\ [-0.7ex]
\hspace{1mm}&
$Q_{3}$ &\hspace{1mm} &  $
\frac{i}{2}\left(a_{1}^{\dag}a_{2}^{\dag} - a_{1}a_{2}\right)
       $ & \hspace{1mm}&
$\frac{1}{2}\pmatrix{I   & 0\cr 0 & I} \sigma_{2}$
\\[2.5ex]
%----------------------------------------------
\hline
\hline\\[-0.4ex]
\end{tabular}
\end{center}
\end{table}
%----------------------------------------------------------------------------------

We can then write the generators of the $Sp(4)$ group as~\cite{bkn19}
\begin{eqnarray}\label{eq11}
&{}& J_{1} = -\frac{1}{2}\pmatrix{0 & I\cr I & 0}\sigma_{2}  , \quad
J_{2} = \frac{i}{2} \pmatrix{0 & -I \cr I & 0} I , \quad
J_{3} = \frac{1}{2}\pmatrix{-I & 0 \cr 0 & I}\sigma_{2} , \nonumber \\[3ex]
&{}& S_{0} = \frac{1}{2}\pmatrix{I   & 0\cr 0 & I} \sigma_{2},
\end{eqnarray}
\noindent and
\begin{eqnarray}\label{323}
&{}& K_{1} = \frac{i}{2}\pmatrix{I  & 0 \cr 0 & -I} \sigma_{1}, \quad
K_{2} = \frac{i}{2} \pmatrix{I & 0 \cr 0 & I} \sigma_{3}, \quad
K_{3} = -\frac{i}{2}\pmatrix{0 & I \cr I & 0}  \sigma_{1}, \nonumber \\[2ex]
&{}& Q_{1} = -\frac{i}{2}\pmatrix{I & 0 \cr 0 & -I}\sigma_{3}, \quad
 Q_{2} = \frac{i}{2}\pmatrix{I & 0 \cr 0 & I}\sigma_{1} , \quad
Q_{3} = \frac{i}{2}\pmatrix{0 &  I \cr I  & 0} \sigma_{3} .
\end{eqnarray}

Among these ten matrices, six of them are in block-diagonal form.  They are
$S_{0}, J_{3}, K_{1}, K_{2} , Q_{1},$ and $Q_{2} .$
In the language of two harmonic oscillators, these generators do not
mix up the first and second oscillators.   There are six of them because
each operator has three generators for its own $Sp(2)$ symmetry.
These generators, together with those in the oscillator representation, are
tabulated in Table~\ref{tab22}.

The off-diagonal matrix $J_{2}$ couples the first and second oscillators
without changing the overall volume of the four-dimensional phase space.
However, in order to construct the closed set of commutation relations,
we need the three additional generators: $J_{1} , K_{3},$ and $Q_{3}.$
The commutation relations given in Eqs.(\ref{313}) are clearly
consequences of Heisenberg's uncertainty relations.

As for the $O(3,2)$ group, the generators are five-by-five matrices, applicable
to $(x, y, z, t, s)$, where $t$ and $s$ are time-like variables.   These matrices
can be written as
\begin{eqnarray}\label{350}
&{}& J_{1} = \pmatrix{0 & 0 & 0 & 0 & 0 \cr 0 & 0 & -i & 0 & 0 \cr
0 & i & 0 & 0 & 0 \cr 0 & 0 & 0 & 0 & 0 \cr 0 & 0  & 0 & 0 & 0 }, \quad
J_{2} = \pmatrix{0 & 0 & i & 0 & 0 \cr 0 & 0 & 0 & 0 & 0 \cr
-i & 0 & 0 & 0 & 0 \cr 0 & 0 & 0 & 0 & 0 \cr 0 & 0  & 0 & 0 & 0 }, \quad
J_{3} = \pmatrix{0 & -i & 0 & 0 & 0 \cr i & 0 & 0 & 0 & 0 \cr
0 & 0 & 0 & 0 & 0 \cr 0 & 0 & 0 & 0 & 0 \cr 0 & 0  & 0 & 0 & 0 } , \nonumber \\[2ex]
  &{}&
K_{1} = \pmatrix{0 & 0 & 0 & i & 0 \cr 0 & 0 & 0 & 0 & 0 \cr
0 & 0 & 0 & 0 & 0 \cr i & 0 & 0 & 0 & 0 \cr 0 & 0  & 0 & 0 & 0 }, \quad
K_{2} = \pmatrix{0 & 0 & 0 & 0 & 0 \cr 0 & 0 & 0 & i & 0 \cr
0 & 0 & 0 & 0 & 0 \cr 0 & i & 0 & 0 & 0 \cr 0 & 0  & 0 & 0 & 0 }, \quad
K_{3} = \pmatrix{0 & 0 & 0 & 0 & 0 \cr 0 & 0 & 0 & 0 & 0 \cr
0 & 0 & 0 & i & 0 \cr 0 & 0 & i & 0 & 0 \cr 0 & 0  & 0 & 0 & 0 } , \nonumber \\[2ex]
&{}&
Q_{1} = \pmatrix{0 & 0 & 0 & 0 & i \cr 0 & 0 & 0 & 0 & 0 \cr
0 & 0 & 0 & 0 & 0 \cr 0 & 0 & 0 & 0 & 0 \cr i & 0  & 0 & 0 & 0 }, \quad
Q_{2} = \pmatrix{0 & 0 & 0 & 0  & 0 \cr 0 & 0 & 0 & 0 & i \cr
0 & 0 & 0 & 0 & 0 \cr 0 & 0 & 0 & 0 & 0 \cr 0 & i  & 0 & 0 & 0 }, \quad
Q_{3} = \pmatrix{0 & 0 & 0 & 0 & 0 \cr 0 & 0 & 0 & 0 & 0 \cr
0 & 0 & 0 & 0 & i \cr 0 & 0 & 0 & 0 & 0 \cr 0 & 0  & i & 0 & 0 } ,  \nonumber \\[2ex]
&{}&
S_{0} = \pmatrix{0 & 0 & 0 & 0 & 0 \cr 0 & 0 & 0 & 0 & 0 \cr
0 & 0 & 0 & 0 & 0 \cr 0 & 0 & 0 & 0 & -i \cr 0 & 0  & 0 & i & 0 } .
\end{eqnarray}

Next, we are interested in eliminating all the elements in the fifth row.
The six generators $J_{i}$ and $K_{i}$ are not affected by this operation, but
$Q_{1}, Q_{2}, Q_{3},$ and $S_{0}$ become
\begin{eqnarray}\label{351}
&{}& P_{1} = \pmatrix{0 & 0 & 0 & 0 & i \cr 0 & 0 & 0 & 0 & 0 \cr
0 & 0 & 0 & 0 & 0 \cr 0 & 0 & 0 & 0 & 0 \cr 0 & 0  & 0 & 0 & 0 }, \quad
P_{2} = \pmatrix{0 & 0 & 0 & 0  & 0 \cr 0 & 0 & 0 & 0 & i \cr
0 & 0 & 0 & 0 & 0 \cr 0 & 0 & 0 & 0 & 0 \cr 0 & 0  & 0 & 0 & 0 }, \quad
P_{3} = \pmatrix{0 & 0 & 0 & 0 & 0 \cr 0 & 0 & 0 & 0 & 0 \cr
0 & 0 & 0 & 0 & i \cr 0 & 0 & 0 & 0 & 0 \cr 0 & 0  & 0 & 0 & 0 } ,\nonumber\\[2ex]
&{}&
P_{0} = \pmatrix{0 & 0 & 0 & 0 & 0 \cr 0 & 0 & 0 & 0 & 0 \cr
0 & 0 & 0 & 0 & 0 \cr 0 & 0 & 0 & 0 & -i \cr 0 & 0  & 0 & 0 & 0 } ,
\end{eqnarray}
respectively. While $J_{i}$ and $K_{i}$ generate Lorentz transformations on the
four dimensional Minkowski space, these $Q_{i}$ and $S_{0}$ in the
form of the $P_i, P_0$ matrices generate translations along the
$x, y, z,$ and $t$ directions respectively.  We shall study this aspect in detail
in Sec.~\ref{contrac}.

%%%%%%%%%%%%%%%%%%%%%%%%%%%%%%%%%%%%%%%%%%%%%%%%%%%%%%%%%%%%%%%%%%%%%%%%%%%%%%%%%%%%%%%%
 \newpage

\section{Contraction of O(3,2) to the Inhomogeneous\\  Lorentz Group}\label{contrac}

We can contract $O(3,2)$ according to the procedure introduced by
In{\"o}n{\"u} and Wigner~\cite{inonu53}.  They introduced the procedure for
transforming the four-dimensional Lorentz group into the three-dimensional
Galilei group.  Here, we shall contract the boost generators belonging to the
time-like $s$ variable, $Q_{i}$ , along with the rotation generator between
the two time-like variables, $S_{0}$.

Here, we illustrate the In{\"o}n{\"u}-Wigner procedure using the concept of
squeeze transformations. For this purpose, let us introduce the squeeze matrix
\begin{equation}\label{335}
C(\epsilon) = \pmatrix{1/\epsilon & 0 & 0 & 0 & 0 \cr
0 & 1/\epsilon & 0 & 0 & 0 \cr 0 & 0 & 1/\epsilon & 0 & 0 \cr
0 & 0 & 0 & 1/\epsilon & 0 \cr 0 & 0  & 0 & 0 &  \epsilon  } .
\end{equation}
This mtrix commutes with $J_{i}$ and $K_{i}$.  The story is different for
$Q_{i}$ and $S_{0}$.

For $Q_{1}$,
\begin{equation} \label{336}
C~Q_{1}~C^{-1} = \pmatrix{0 & 0 & 0 & 0 & i/\epsilon^{2} \cr
   0 & 0 & 0 & 0 & 0 \cr 0 & 0 & 0 & 0 & 0 \cr 0 & 0 & 0 & 0 & 0 \cr
   i\epsilon^{2}  & 0  & 0 & 0 & 0 },
\end{equation}
which, in the limit of small $\epsilon$, becomes
\begin{equation}
Q'_{1} = \pmatrix{0 & 0 & 0 & 0 & i/\epsilon^{2} \cr 0 & 0 & 0 & 0 & 0 \cr
0 & 0 & 0 & 0 & 0 \cr 0 & 0 & 0 & 0 & 0 \cr 0  & 0  & 0 & 0 & 0 }  .
\end{equation}
We then make the inverse squeeze transformation:
\begin{equation}\label{339}
   C^{-1}~Q'_{1}~C = \pmatrix{0 & 0 & 0 & 0 & i \cr 0 & 0 & 0 & 0 & 0 \cr
0 & 0 & 0 & 0 & 0 \cr 0 & 0 & 0 & 0 & 0 \cr 0  & 0  & 0 & 0 & 0 }.
\end{equation}

%----------------------------------------------------------------------
%----------------------------------------------------------------------

Thus, we can write this contraction procedure as
\begin{equation}
P_{1} = \lim_{\epsilon\rightarrow 0 }\left(\epsilon^{2}~C~Q_{1}~C^{-1}\right),
\end{equation}
where the explicit five-by-five matrix is given in Eq.(\ref{351}).
Likewise
\begin{equation}
   P_{2} = \lim_{\epsilon\rightarrow 0 }\left(\epsilon^{2}~C~Q_{2}~C^{-1}\right),  \quad
   P_{3} = \lim_{\epsilon\rightarrow 0 }\left(\epsilon^{2}~C~Q_{3}~C^{-1}\right),  \quad
   P_{0} = \lim_{\epsilon\rightarrow 0 }\left(\epsilon^{2}~C~S_{0}~C^{-1}\right).
\end{equation}
These four contracted generators lead to the five-by-five transformation matrix, as can be
seen from
\begin{equation}\label{349}
\exp\left\{-i\left(aP_{1}+ bP_{2} + cP_{3} + dP_{0}\right)\right\}
\end{equation}
performing translations in the four-dimensional Minkowski space:
\begin{equation}\label{351a}
     \pmatrix{1 & 0 & 0 & 0 & a \cr 0 & 1 & 0 & 0 & b \cr
0 & 0 & 1 & 0 & c \cr 0 & 0 & 0 & 1 & -d \cr 0 & 0  & 0 & 0 & 1 }
\pmatrix{x \cr y \cr z \cr t \cr 1 } =
\pmatrix{x + a \cr y + b \cr z + c \cr t - d \cr 1 } .
\end{equation}

In this way, the space-like directions are translated and the time-like $t$
component is shortened by an amount $d$. This means the group $O(3,2)$
derivable from the Heisenberg's
uncertainty relations becomes the inhomogeneous Lorentz group governing
the Poincar\'e symmetry for quantum mechanics and quantum field theory.
These matrices correspond to the differential operators
\begin{equation}
 P_{x} = -i\frac{\partial}{\partial x}, \quad
 P_{y} = -i\frac{\partial}{\partial y}, \quad
 P_{z} = -i\frac{\partial}{\partial z}, \quad
 P_{0} = i\frac{\partial}{\partial t },
\end{equation}
respectively. These translation generators correspond to the Lorentz-covariant
four-momentum variable with
\begin{equation}
     p_{1}^{2} + p_{2}^{2} + p_{3}^{2} - p_{0}^{2}  = \mbox{constant} .
\end{equation}
This energy-momentum relation is widely known as Einstein's $E = mc^{2}$.

%%%%%%%%%%%%%%%%%%%%%%%%%%%%%%%%%%%%%%%%%%%%%%%%%%%%%%%%%%%%%%%%%%%%%%%%%%%%%%%%%%%%%%%%

\newpage

\section*{Concluding Remarks}

According to Dirac~\cite{dir49}, the problem of finding a
Lorentz-covariant quantum mechanics reduces to the problem of
finding a representation of the inhomogeneous Lorentz group.  Again,
according to Dirac~\cite{dir63}, it is possible to construct the Lie
algebra of the group $O(3,2)$ starting from two oscillators.  We
have shown in our earlier paper~\cite{bkn19} that this $O(3,2)$
group can be contracted to the inhomogeneous Lorentz group
according to the group contraction procedure introduced by
In{\"o}n{\"u} and Wigner~\cite{inonu53}.

In this paper, we noted first that the symmetry of a single oscillator
is generated by three generators.  Two independent oscillators
thus have six generators.  We have shown that there are four additional
generators needed for the coupling of the two oscillators.  Thus there
are ten generators.  These ten generators can then be linearly combined
to produce ten generators which were spelled out in Dirac's 1963 paper.

For the two-oscillator system, there are four step-up and step-down
operators.  There are therefore sixteen quadratic forms~\cite{hkn95jmp}.
Among those, only ten of them are in Dirac's 1963 paper~\cite{dir63}.
Why ten?  Dirac needed those ten to construct the Lie algebra for the
$O(3,2)$ group.  At the end of the same paper, he stated that this Lie
algebra is the same as that for the $Sp(4)$ group, which preserves the
minimum uncertainty for each oscillator.

In this paper, we started with the block-diagonal matrix given in
Eq.(\ref{m03}) for two totally independent oscillators with six
independent generators.  We then added one two-by-two Hermitian matrix
of Eq.(\ref{m05}) with four generators for the off-diagonal blocks.
The result is the four-by-four Hermitian matrix given in Eq.(\ref{m07}).
This four-by-four Hermitian matrix has ten independent operators which
can be linearly combined to the ten operators chosen by Dirac.  Thus,
in this paper, we have shown how the two-oscillators are coupled, and
how this coupling introduces additional symmetries.

Paul A. M. Dirac made his life-long efforts to make quantum mechanics
consistent with special relativity, starting from 1927~\cite{dir27}.
While we exploited the contents of his paper published in 1963~\cite{dir63},
it is of interest to review his earlier efforts.  In his earlier papers,
Dirac started with quantum mechanics and special relativity as two
different branches of science based on two different mathematical bases.

In this paper, based on Dirac's two papers~\cite{dir49,dir63}, we concluded
that both quantum mechanics and special relativity can be derived from
the same mathematical base. A brief review of Dirac's earlier efforts is given
in the Appendix.

\newpage

\section*{Appendix}

As we all know, quantum mechanics and special relativity were developed
along two separate routes.  As early as 1927, Dirac was interested in
understanding whether these two scientific disciplines are compatible
with each other.   In his paper of 1927~\cite{dir27}, Dirac noted the
the existence of the time-energy uncertainty relation without excitations.
He called this the ``c-number'' time-energy uncertainty relation.
Dirac pointed out that the space-time asymmetry makes it difficult
to construct the uncertainty relation in the Lorentz-covariant world.

%----------------------------------------------------------------------
\begin{figure}[thb]
\centerline{\includegraphics[width=12cm]{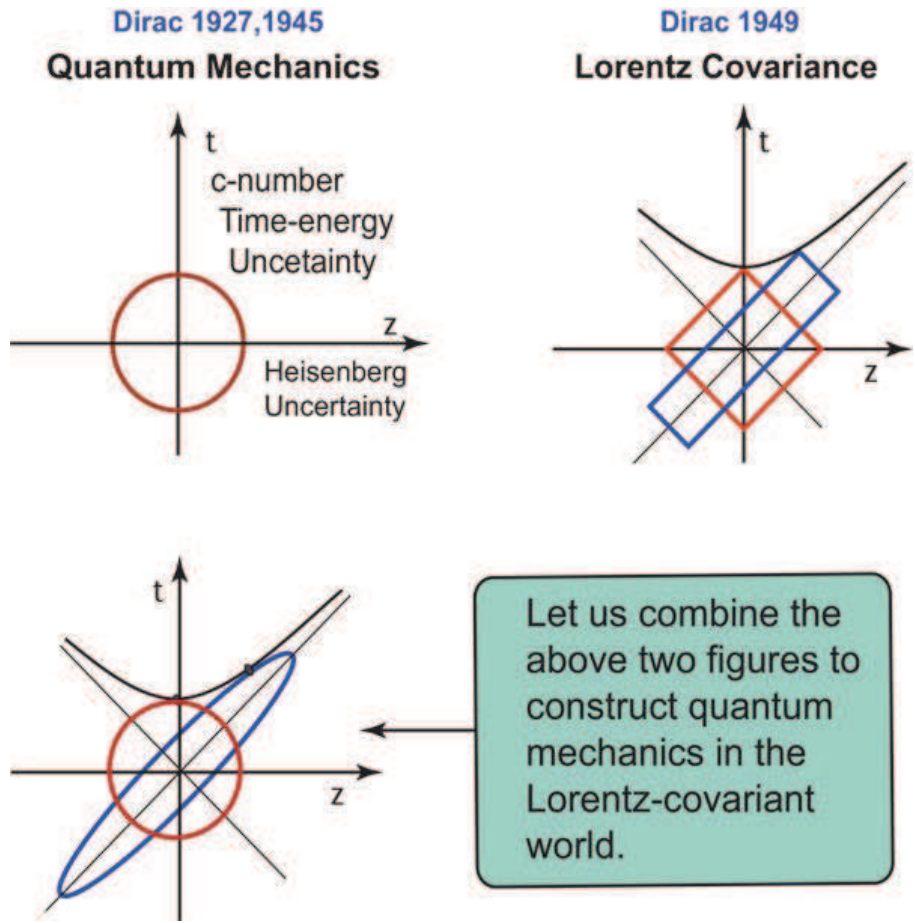}}
\caption{Dirac's three papers. His 1927 and 1945 papers can be described
by a circle in the longitudinal space-like  and time-like coordinate.
Dirac introduced the light-cone coordinate system in 1959.  In this
system, the Lorentz boost is a squeeze transformation.  It is then
natural to synthesize these two figures to a squeezed circle or an
ellipse.  Figure~\ref{parton33} will illustrate how this elliptic
squeeze manifests itself in the real world.}\label{dqm88}
\end{figure}

%----------------------------------------------------------------------
In 1945, Dirac considered the four-dimensional harmonic oscillator wave
functions applicable to the four-dimensional space and time.  In so
doing, Dirac was considering localized bound states.  The space and time
variables in his case are the separations between two constituents, like
the proton and electron in the hydrogen atom.

It was shown later that Dirac's concern about the c-number time-energy
uncertainty is not necessary in view of the fact that a massive particle
at rest has only three space-like dimensions~\cite{kno79jmp}.  According
to Wigner~\cite{wig39}, the little group for the massive particle is
isomorphic to $O(3)$~\cite{wig39}.   With this understanding, we can use
a circle in the $z~t$ plane as shown in Fig.~\ref{dqm88}, where $z$ and
$t$ are longitudinal and time separations respectively.

In his 1949 paper~\cite{dir45}, Dirac introduced the light-cone coordinate
system which tells us that the Lorentz boost is a squeeze transformation.
This aspect is also illustrated in Fig.~\ref{dqm88}.  It is then not
difficult to see how the circle looks to a moving observer.

%-------------------------------------------------------------------------------
\begin{figure}%[thb]
\centerline{\includegraphics[width=16cm]{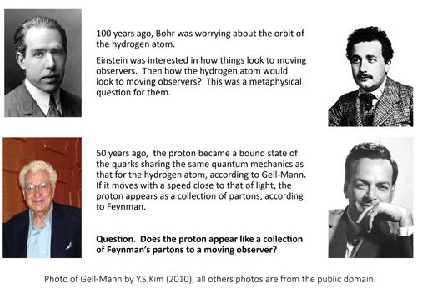}}
\caption{The Bohr-Einstein issue is 100 years old.  Fifty years later,
it became the quark-parton puzzle, based on observations made in high-energy
laboratories.}\label{einbo}
\end{figure}
%-------------------------------------------------------------------------------

Next question is whether this elliptic squeeze has anything to do with
the real world.  One hundred years ago, Niels Bohr and Albert Einstein
met occasionally to discuss physics.  Their interests were different.
Bohr was worrying about the electron orbit in the hydrogen atom.
Einstein was interested in how things look to moving observers.  Then
the question arises.  How would the hydrogen atom look to a moving
observer?  This was a metaphysical issue during the period of Bohr
and Einstein, because there were no hydrogen atoms moving fast enough
to exhibit this Einstein effect.

Fifty years later, the physics world was able to produce many protons from
particle accelerators.  In 1964~\cite{gell64}, Gell-Mann observed that the
proton is a bound state of the more fundamental particles called ``quarks''
according to the quantum mechanics applicable also to the hydrogen atom.

\begin{figure} [thb]
\centerline{\includegraphics[width=16cm]{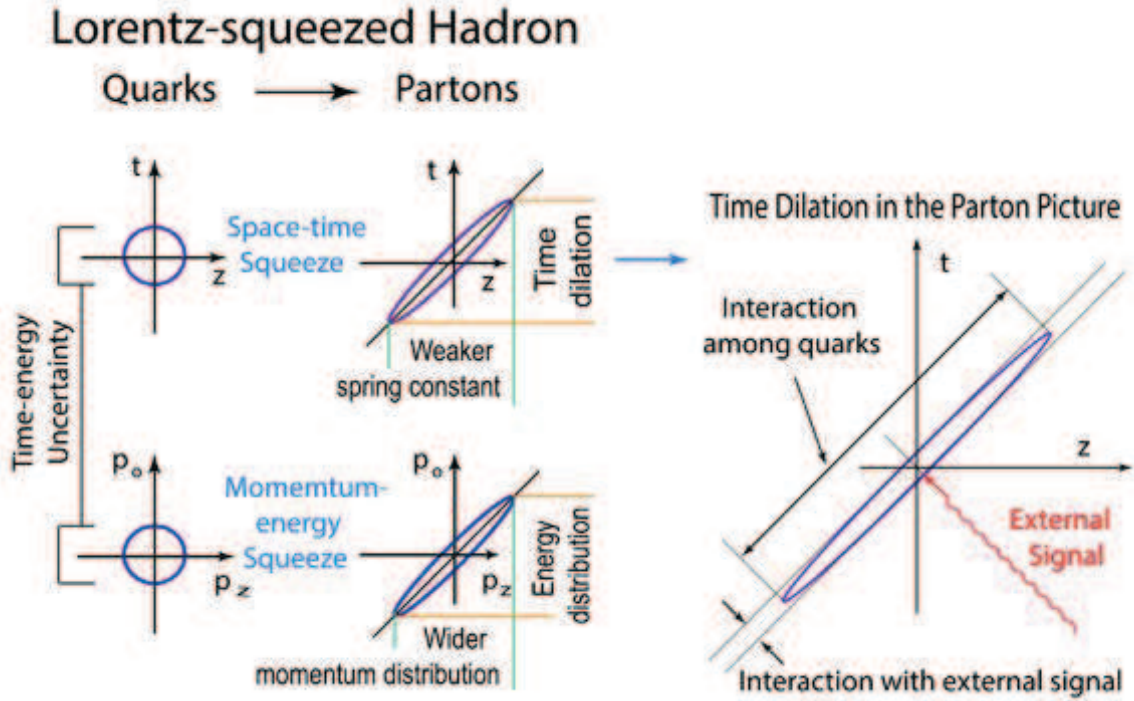}}
\caption{In the harmonic-oscillator regime, the momentum-energy wave function
takes the same mathematical form as that of the space-time wave functions.
This figure shows that the quark model and the parton model are two different
aspects of one Lorentz-covariant entity.
In 1969~\cite{fey69}, Feynman observed that the fast-moving proton appears
as a collection of a large number of light-like partons with a wide-spread
momentum distribution, and short interaction time with the external signal.
This figure is a graphical illustration of the 1977 paper by Kim and
Noz~\cite{kn77}.   This figure is from a recent book by the present
authors~\cite{bks15iop}.}\label{parton33}
\end{figure}

However, according to Feynman~\cite{fey69,bj69}, when the proton moves
very fast, it appears as a collection of a large number of free-moving
light-like partons with a wide-spread momentum distribution, as described
in Fig.~\ref{parton33}.  Feynman's parton picture was entirely based on
what we observe in laboratories.

Unlike the hydrogen atom, the proton can become accelerated, and its speed
could be very close to that of light.  Thus the Bohr-Einstein issue became
the Gell-Mann-Feynman issue, as illustrated in Fig.~\ref{einbo}.  The question
is whether Gell-Mann's quark model and Feynman's parton picture are two different
aspects of one Lorentz-covariant entity.  This question was addressed by Kim and
Noz 1977~\cite{kn77} and was explained in detail by the present authors with
a graphical illustration  given in Fig.~\ref{parton33}.

%-------------------------------------------------------------------------------
%----------------------------------------------------------------------

\newpage

%%%%%%%%%%%%%%%%%%%%%%%%%%%%%%%%%%%%%%%%%%%%%%%%%%%%%%%%%%%%%%%%%%%%%%%%%%%%%%%%%%%%%%%%


\begin{thebibliography}{99}

\bibitem{dir49}
Dirac, P. A. M.
  Forms of Relativistic Dynamics.
  {\em Rev. Mod. Phys.} {\bf 1949} {\em 21}, 392 - 399.



\bibitem{dir63}
Dirac, P. A. M.
  A Remarkable Representation of the 3 + 2 de Sitter Group.
  {\em J. Math. Phys.}  {\bf 1963} {\em 4}, 901 - 909.

\bibitem{inonu53}
In{\"o}n{\"u}, E.; Wigner, E. P.
  On the Contraction of Groups and their Representations.
  {\em Proc. Natl. Acad. Sci. (U.S.)}
  {\bf 1953} {\em 39}, 510 - 524.


\bibitem{dir27}
Dirac, P. A. M.
  The Quantum Theory of the Emission and Absorption of Radiation
  {\em Proc. Roy. Soc. (London)} {\bf 1927} {\em A114} 243 - 265.


\bibitem{hkn88}
Han, D.; Kim, Y. S.; Noz, M.E.
Linear canonical transformations of coherent and squeezed
states in the Wigner phase space.
{\em Phys. Rev. A} {\bf 1988} {\em 37}, 807 - 814.

\bibitem{kiwi90ajp}
Kim, Y. S.; Wigner, E. P.
Canonical transformation in quantum mechanics.
{\em Am. J. Phys.} {\bf 1990}, {\em 58}, 439 - 447.

\bibitem{knp91}
Kim, Y. S.; Noz, M. E.
  {\em Phase Space Picture of  Quantum Mechanics};
  World Scientific Publishing Company: Singapore, 1991.


\bibitem{dodo03}
Dodonov, V. V.; Man'ko V. I.
{\em Theory of Nonclassical States of Light};
Taylor \& Francis: London \& New York, 2003.


\bibitem{hkn95jmp}
Han, D.; Kim, Y. S.; Noz, M. E.
  $O(3,3)$-like Symmetries of Coupled Harmonic Oscillators.
  {\em J. Math. Phys.} {\bf 1995} {\em 36}, 3940 - 3954.

\bibitem{abra78}
Abraham, R.; Marsden, J. E.
  {\em Foundations of Mechanics 2nd ed.}
  Benjamin Cummings: Reading, Massachusetts, 1978.

\bibitem{goldstein80}
Goldstein, H.
  {\em Classical Mechanics.  2nd ed.}
  Addison-Wesley: Reading, Massachusetts, 1980.

\bibitem{bkn19}
Ba{\c s}kal, S.; Kim, Y. S.; Noz, M. E.
 Poincar\'e Symmetry from Heisenberg's Uncertainty Relations.
 {\em Symmetry} {\bf 2019} {\em 11}, (3) 49:1 - 9.


\bibitem{kno79jmp}
Kim, Y. S.; Noz, M. E.; Oh, S. H.
  Representations of the Poincar\'e group for relativistic extended hadrons
  {\em J. Math. Phys.} {\bf 1979} {\em 20} 1341 - 1344.

\bibitem{wig39}
Wigner, E.
 On unitary representations of the inhomogeneous Lorentz group
 {\em Ann. Math.} {\bf 1939}, {\em 40}, 149 - 204.


\bibitem{dir45}
Dirac, P. A. M.
  Unitary Representations of the Lorentz Group
  {\em Proc. Roy. Soc. (London)} {\bf 1945} {\em A183} 284 - 295.


\bibitem{gell64}
Gell-Mann, M.
  A Schematic Model of Baryons and Mesons
  {\em Phys. Lett.} {\bf 1964} {\em 8}, 214 - 215.

\bibitem{fey69}
Feynman, R. P.
  Very High-Energy Collisions of Hadrons
  {\em Phys. Rev. Lett.} {\bf 1969} {\em 23} 1415 - 1417.

\bibitem{bj69}
Bjorken, J. D.; Paschos, E. A.
  Electron-Proton and $\gamma$-Proton Scattering and the Structure
  of the Nucleon
  {\em Phys. Rev.} {\bf 1969} {\bf 185} 1975 - 1982.


\bibitem{kn77}
Kim, Y. S.; Noz, M. E.
  Covariant harmonic oscillators and the parton picture
  {\em Phys. Rev. D} {\bf 1977} {\em 15} 335 - 338.

\bibitem{bks15iop}
Ba{\c s}kal, S; Kim. Y. S.; E. Noz, M. E.
 {\em Physics of the Lorentz Group, IOP Concise Physics}
  Morgan \& Claypool Publisher, San Rafael, California, U.S.A.
  and IOP Publishing, Bristol, UK., 2015.






\end{thebibliography}
\end{document}